\documentclass{elsarticle}
\usepackage{graphicx}
\usepackage{amsmath}
\usepackage{amssymb}

\usepackage{txfonts}
\def\frac#1#2{{#1 \over #2}}

\def\half{\ifinner {\scriptstyle {1 \over 2}}
   \else {1 \over 2} \fi}

\def\ket#1{\vert#1\rangle}              

\def\simge{\mathrel{%
   \rlap{\raise 0.511ex \hbox{$>$}}{\lower 0.511ex \hbox{$\sim$}}}}
\def\simle{\mathrel{
   \rlap{\raise 0.511ex \hbox{$<$}}{\lower 0.511ex \hbox{$\sim$}}}}

\def\slashchar#1{\setbox0=\hbox{$#1$}           
   \dimen0=\wd0                                 
   \setbox1=\hbox{/} \dimen1=\wd1               
   \ifdim\dimen0>\dimen1                        
      \rlap{\hbox to \dimen0{\hfil/\hfil}}      
      #1                                        
   \else                                        
      \rlap{\hbox to \dimen1{\hfil$#1$\hfil}}   
      /                                         
   \fi}                                         %

\def\subrightarrow#1{
  \setbox0=\hbox{
    $\displaystyle\mathop{}
    \limits_{#1}$}
  \dimen0=\wd0
  \advance \dimen0 by .5em
  \mathrel{
    \mathop{\hbox to \dimen0{\rightarrowfill}}
       \limits_{#1}}}                           

\usepackage{amssymb}

\usepackage{makeidx}

\usepackage{bbold} 

\def\simge{\mathrel{%
    \rlap{\raise 0.511ex \hbox{$>$}}{\lower 0.511ex \hbox{$\sim$}}}}
\def\simle{\mathrel{
    \rlap{\raise 0.511ex \hbox{$<$}}{\lower 0.511ex \hbox{$\sim$}}}}

\def\x{{\boldsymbol x}}
\def\y{{\boldsymbol y}}

\newcommand \beq{\begin{eqnarray}}
\newcommand \eeq{\end{eqnarray}}
\newcommand{\del}{\partial}

\journal{Nuclear Physics A}
\begin{document}


\begin{frontmatter}

\title{Many-Body Physics: \\
Collective fermionic excitations in quark-gluon plasmas and cold atom systems}

\author{Jean-Paul Blaizot$^1$}

\address{$^1$Institut de Physique Th\'eorique, CNRS/URA2306, \\CEA-Saclay, 91191 Gif-sur-Yvette, France}


\begin{abstract}
\noindent
In this talk I discuss collective excitations that carry fermion quantum numbers. Such excitations occur in the  quark-gluon plasma and can also be produced in cold atom systems under special conditions. 
\end{abstract}
\end{frontmatter}


\section{Introduction}

My first memories of Gerry bring me back to my very first year in physics. Indeed I first  met Gerry as I was just starting my PhD. He used to visit Saclay regularly then, and everyone in the nuclear theory group was eager to show him his/her last results, and get his opinion. Certainly, like everybody else, I was also under the charm of his exceptional personality, and was indeed surprised that he would pay any attention to me and show curiosity for what I was doing. At the same time, I must confess that it took me some time to truly appreciate Gerry's way of doing physics. My academic training, and the particular environment of the IPhT, where mathematical rigor and elegance were somewhat more praised than physical intuition, did not help me there. However, with time, I had many occasions to be impressed by Gerry's intuition, his art of formulating good questions and distributing interesting problems to his (many) young collaborators, and his ability of getting deep physical insight into a complicated problem through a simple, but to the point, calculation.  

When I first met Gerry, his book, {\em Many-Body Physics}, had just appeared\cite{MBPbs}. This book has accompanied me during my first years of research. I learned from it, asides from the many discussions I have had with Gerry, about the basics of ``traditional'' many body physics: Fermi liquid theory, collective excitations in a variety of systems and the Random Phase Approximation, Brueckner theory, etc. It is with all this in mind that I have been preparing this talk.
 I have chosen to focus on  collective  excitations of a somewhat peculiar nature,  since they carry fermionic degrees of freedom (most collective modes we are familiar with have bosonic character). I shall start with the quark gluon plasma, where fermionic excitations have been identified (theoretically !) long ago. I shall discuss the very long wavelength modes that can be interpreted as Goldstone excitations associated to a broken (approximate) supersymmetry. Then I shall move to cold atom systems, where efforts are being made to produce analogous phenomena. 
 
 \section{Quark-gluon plasma}

  Let us then consider   a  quark-gluon plasma. I shall assume the temperature sufficiently high for the coupling constant $g$ to be small.  (I therefore leave aside interesting issues related to so-called strongly coupled quark-gluon plasma, the AdS/CFT correspondence, etc.) The basic degrees of freedom in such a system are quarks and gluons quasiparticles. From the point of view of many body physics, such systems are interesting because the long range interactions favor the emergence of collective phenomena. 
  
In a weakly coupled quark-gluon plasma, one can identify a hierarchy of scales that are proportional to the temperature, multiplied by various powers of $g$.  Although these scales may not be well separated when the coupling constant is not sufficiently small, the physical description based on this hierarchy is physically consistent. This is therefore a useful organization principle. 
 
 The existence of such a hierarchy can be understood simply, by realizing that  the expansion parameter in a field theory at finite temperature depends not only on the strength of the coupling, but also on the magnitude of the relevant thermal fluctuations. Consider for instance a scalar field theory, with a $g^2\phi^4$ interaction. The thermal fluctuations are given by 
\beq\label{fluctuations}
\langle \phi^2\rangle= \int\frac{d^3 k}{(2\pi)^3}\frac{n_k}{k}, \qquad n_k=\frac{1}{{\rm e}^{k/T}-1}.
\eeq
The integral in (\ref{fluctuations}) is dominated by the largest values of $k$. We estimate it with an upper cut-off $\kappa$ and refer to the corresponding value as to ``the contribution of the fluctuations at scale $\kappa$'', and denote it by $\langle \phi^2\rangle_\kappa$. In the same spirit, we approximate the kinetic energy as $ \langle ( \del\phi)^2\rangle_\kappa \approx \kappa^2 \langle \phi^2\rangle_\kappa$. Taking furthermore $ \langle\phi^4\rangle_\kappa \approx \langle \phi^2\rangle_\kappa^2$, one gets as expansion parameter (ratio of potential to kinetic energies)
\beq
\gamma_\kappa=\frac{g^2 \langle \phi^2\rangle_\kappa}{\kappa^2}\simeq \frac{g^2T}{\kappa}, \qquad \langle \phi^2\rangle_T\sim \kappa T,
\eeq
where the last equality is valid for $\kappa\lesssim T$ (then $n_k\sim T/k$).
The fluctuations that dominate the energy density at weak coupling correspond to the plasma particles and have momenta $k\sim T$. For these ``hard'' fluctuations, 
$\kappa\sim T$, $ \langle \phi^2\rangle_T\sim T^2$, and $ \gamma_T\sim g^2$.
At this scale,  perturbation
theory works as well as at zero temperature (with expansion parameter $\sim g^2$, or rather $\alpha=g^2/4\pi$). 

The next ``natural'' scale, commonly referred to as the ``soft scale'', corresponds to $\kappa\sim gT$, for which $ \gamma_{gT}\sim g$.
 The expansion parameter  remains small if $g$ is small, but perturbation theory (for the soft modes)  is now an expansion in powers of $g$ rather than $g^2$: it is therefore less rapidly ``convergent''. 
Another phenomenon occurs at the scale $gT$, important for the present discussion. While the expansion parameter $ \gamma_{gT}$ that controls the self-interactions of the soft fluctuations is small, the coupling between the soft modes and the thermal fluctuations at scale $T$ is not: indeed $g^2\langle \phi^2\rangle_T\sim (gT)^2$. Thus the dynamics of soft modes is non-perturbatively renormalized by their coupling to hard modes. This particular coupling is encompassed by the so-called ``hard thermal loops'' (HTL) \cite{HTL}. 

Finally, there is yet another scale, the ``ultra-soft scale'' $\kappa\sim  g^2 T$, at which perturbation theory completely breaks down since $ \gamma_{g^2T}\sim 1$:
ultra-soft  fluctuations remain strongly coupled for arbitrarily small couplings. Of course, this situation does not  occur for a scalar field since the thermal mass $\sim gT$  renders the contribution of the  $g^2T$ fluctuations  negligible.  However this situation is met in QCD for the long wavelength, unscreened, magnetic
fluctuations. 

We shall be interested here in the collective phenomena that occur at the scale $gT$ and below (although I should emphasize that the ultra-soft excitations that I shall discuss do not involve the unsecreend magnetic modes that I just referred to). These excitations can be understood from the properties of the fermion self-energy displayed in Fig.~\ref{self-energy}.\\

\begin{figure}
\begin{center}\label{self-energy}
\includegraphics[scale=0.40]{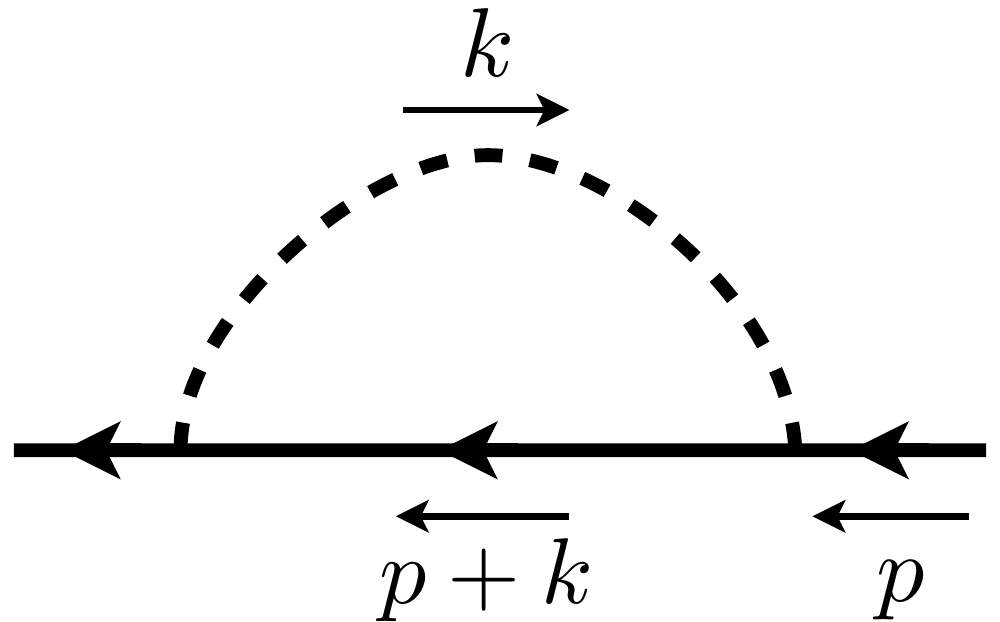}
\caption{The self-energy of quark (full line) coupled to a gluon (dashed line) to order $g^2$. The external momentum is soft ($p\sim gT$) or ultrasoft ($p\sim g^2T$), while the loop momentum is hard ($k\sim T$).\label{self-energy}}
\end{center}
\end{figure}

\noindent{\it The spectrum of soft fermionic excitations}\\

\begin{figure}[h]
\begin{center}
\includegraphics[scale=0.5]{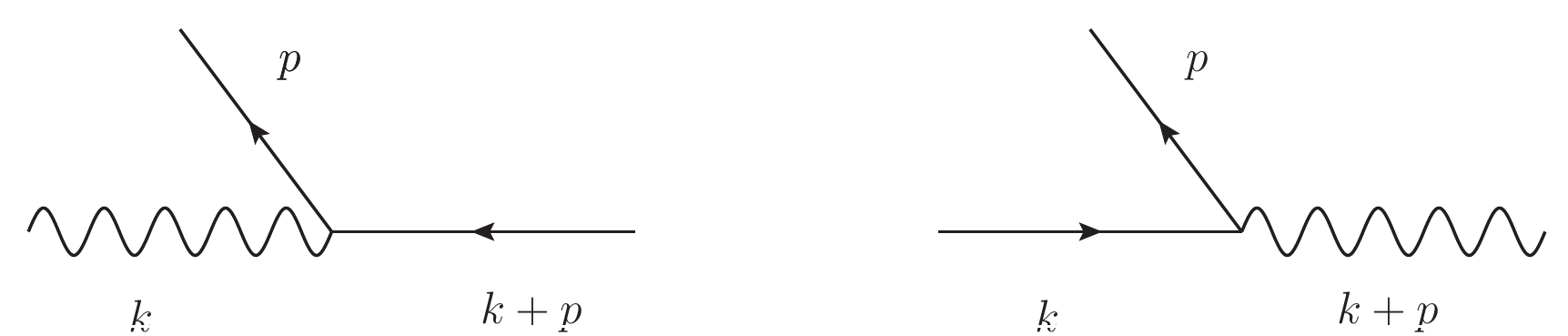}
\caption{~Dominant physical processes contributing to the fermion self-energy. Note the change of nature of the hard ($k\sim T$) particle (fermion to boson and vice versa) in the interaction with the soft fermionic field carrying momentum $p\sim gT$.\label{processes}}
\end{center}
\end{figure}

The retarded self-energy at high temperature is given by 
\beq
\label{eq:sigma-mu}
\Sigma^R(p)\approx g^2\int\frac{d^4k}{(2\pi)^4} L(k) \frac{\slashchar{k}}{\delta m^2+2k\cdot p}.
\eeq
where 
$L(k)\equiv 2\pi\,{\rm sgn}(k^0)\,\delta(k^2)\,(n_f(k^0)+n_b(k^0))$, with $n_f$ and $n_b$ fermion and boson statistical factors, and I have neglected (small) damping terms.  
The momentum integral is dominated by hard momenta ($k\sim T$) of on-shell quasiparticles, and the term $\delta m^2=m_b^2-m_f^2\sim g^2T^2$ is the difference of the squares of the so-called asymptotic thermal masses. The processes involved in the self-energy are 
virtual transitions where hard particles scatter on the soft one, 
with little deflection, and a
possible change in their quantum numbers (see Fig.~\ref{processes}).
 For example, a hard quark with momentum $k\sim T$ can annihilate on a soft
antiquark with momentum $p\sim gT$ 
 and ``turn  into'' a hard gluon (right diagram in Fig.~\ref{processes}). Aside from the changes in quantum
numbers, such processes are reminiscent of the familiar
  Landau damping processes.  We see here emerging an interesting feature, with the soft (or ultra-soft) and hard degrees of freedom playing distinct roles. 
The collective excitations are associated with long wavelength, low frequency, oscillations of the soft degrees of freedom that can be represented by average fields, while the hard, particle,  degrees of freedom  account for the polarization of the medium by the soft fields \cite{Blaizot:2001nr}\footnote{Such oscillating fields involving changes in the number of particles find analogs in nuclear physics with the pairing vibrations \cite{BM}}. This polarization involves here a change in the nature of the hard particles, fermions turning into bosons, and vice versa, reflecting a ``supersymmetric'' behavior that shall be discussed further later.

 \begin{figure}[h]
\begin{center}
\includegraphics[scale=0.65]{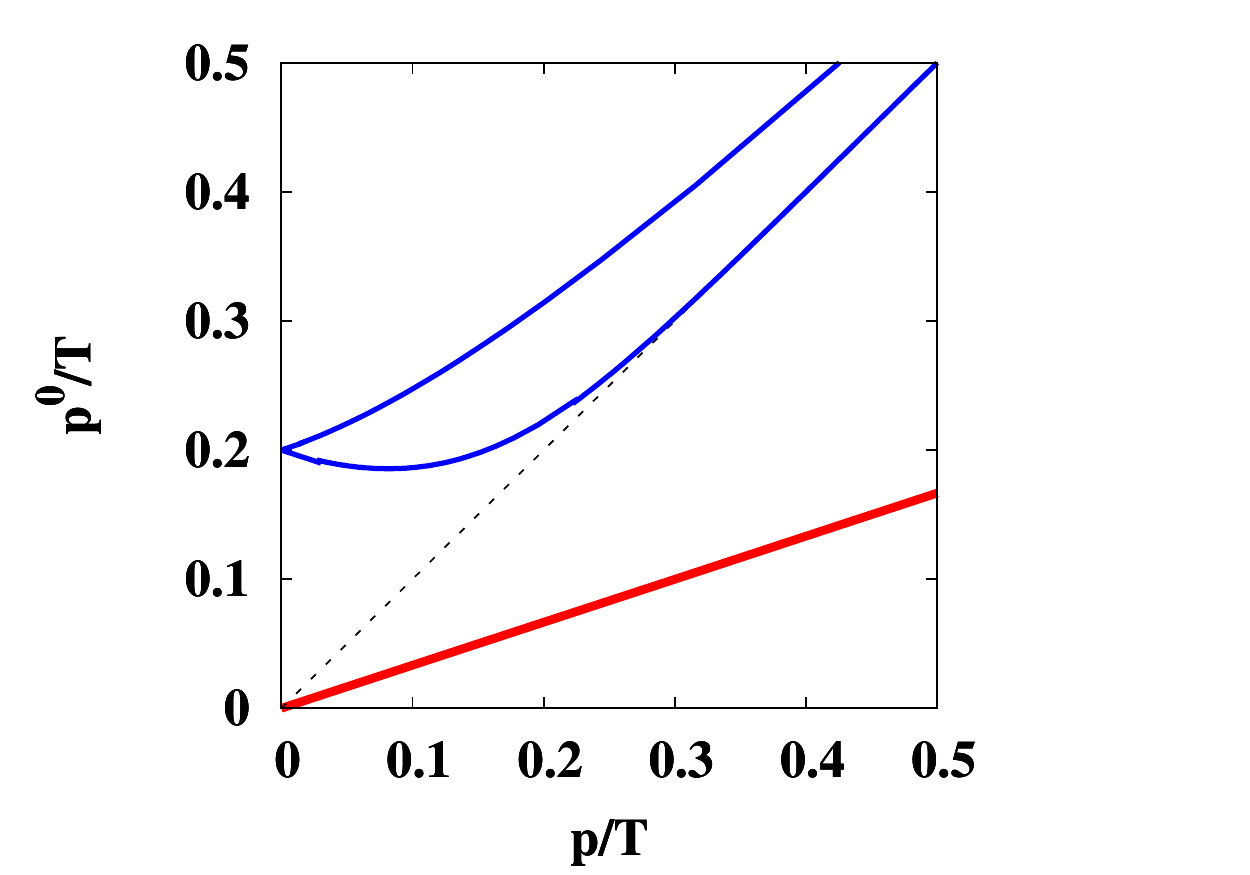}
\caption{The dispersion relation of soft and ultra-soft excitations with positive energies (the spectrum of negative energies is the mirror image of this one) at $T\neq 0$, $\mu=0$, $g=0.8$. 
The upper two branches (solid line, blue) correspond to the plasmino, with the upper branch going into the normal fermion state at high momentum, and the lower branch disappearing from the spectrum when $p\gg gT$.
The lower branch (thick solid line, red) corresponds to the ultra-soft fermionic excitation.
The lower two branches of the spectrum shown in this figure carry quantum numbers that are normally attributed to antiparticles. \label{fermion-spectrum} }
\end{center}
\end{figure}

 The resulting dispersion relations of the corresponding modes are given in Fig.~\ref{fermion-spectrum}.
Let me discuss the main features of  this spectrum, starting from the momentum region of order $gT$. The corresponding mode is called the plasmino, and it is well described by the HTL approximation which corresponds to neglecting  $\delta m^2$ in the denominator in Eq.~(\ref{eq:sigma-mu}), which is legitimate when the external momentum is $\sim gT$ (recall that $k\sim T$). 
A remarkable feature of this spectrum is the split dispersion relation of the plasmino.This can be understood from simple considerations, as I now explain.

The HTL self-energy at very small (on scale $gT$) momentum is given by 
\beq
\Sigma(\omega)=\frac{\omega_0^2}{\omega}\gamma_0,
\eeq
where $\gamma_0$ is a Dirac matrix. I have set $\omega=p_0$ and $\omega_0\sim gT$ is the analog of the plasma frequency. This particular behavior follows immediately from the fact that the processes in Fig.~\ref{processes} contribute an imaginary part in
the region   
 $\left\vert\omega\right\vert <p$, which goes over to a delta function of $\omega$ when $p\to 0$. 
The spectrum at $p=0$ is then obtained from the poles of
\beq
G(\omega)=\frac{\gamma_0}{-\omega+\omega_0^2/\omega}.
\eeq
For $\gamma_0=1$, there are two poles, $\omega_\pm=\pm \omega_0$. There exist
symmetrical poles, at $-\omega_\pm$, corresponding to $\gamma_0=-1$. The states at
$\pm \omega_0$ are therefore doubly degenerate. This degeneracy is removed by a small
fermion mass, or a  finite momentum. This is why, at finite momentum, 
the fermion
dispersion relation at positive frequency appears to be split\cite{Blaizot:1993bb}. 

\begin{figure}[h]
\begin{center}
\includegraphics[scale=0.65]{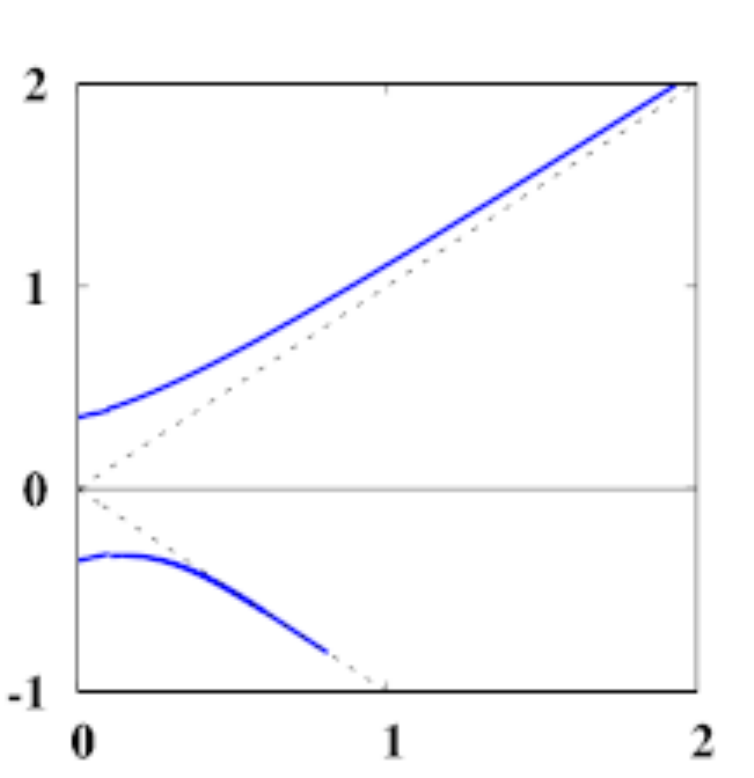}
\caption{ The origin of the split fermion dispersion relation, with $\omega$ on the vertical axis and $p$ on the horizontal one. The coupling of the fermion state corresponding to $\gamma_0>0$ with the continuum of Landau damping processes (concentrated in the region $-p\le \omega\le p$) splits the single particle strength in the two branches displayed in this figure. The lower branch seen in Fig.1 at $\omega>0$ is the mirror image of the negative $\omega$ branch seen here, and corresponds to a negative energy state ($\gamma_0<0$). Hence the unnatural quantum numbers of the plasmino.}
\end{center}
\end{figure}

 The origin of the splitting of, say, the $\gamma_0=1$ state, is  the
coupling of the fermion with a large number of quasi degenerate states. This
mechanism is quite general, and can be understood with the help of a ``schematic
model'', similar to the one Gerry used to describe giant resonances in nuclei \cite{Brown_schematic}. In this model, a quantum state
$\ket{0}$, with energy $E_0$,  couples with equal strength $V$ to a large number $N$
of states $\ket{i}$, uniformly distributed throughout the energy interval
$[-\Delta E/2,\Delta E/2]$ ($\Delta E/2\simle E_0$). In the calculation
performed above,
$\Delta E=2p$, corresponding to the phase space of the Landau damping
processes, and
$V\sqrt{N}\sim gT$. The hamiltonian is:
\beq
H=\left(
\begin{array}{ccccc}
E_0 & V   & \cdot   & \cdot & V\\
V   & E_1 & 0 & \cdot & 0\\
\cdot & 0 & E_2 &\cdot &\cdot\\
\cdot &\cdot & \cdot & \cdot &\cdot\\
V & \cdot &\cdot &\cdot & E_N
\end{array}\right),
\eeq
and the equation determining the modes reads
\beq\label{schematic}
\omega-E_0=\sum_i \frac{V^2}{\omega-E_i}\approx \frac{NV^2}{\Delta E}
\ln\left|\frac{\omega+\Delta E/2}{\omega-\Delta E/2}\right|.
\eeq
Assuming $\omega$ large compared to $\Delta E$, one can expand the logarithm
and get:
\beq
\omega-E_0\approx\frac{NV^2}{\omega}.
\eeq
One can now let $E_0\to 0$ (in this schematic model, $E_0$ plays the role of a small
mass term), and recover the  two solutions at $\omega=\pm\sqrt{NV^2}$ which correspond to the two
poles at $\omega_\pm=\pm M$. In the limit considered, these two solutions
 share the same single particle strength, i.e. 1/2. Note the presence of the factor $N$ in the energy shift: it reflects the collective nature of the phenomenon (in contrast, all the other states are non collective, and their energies $E_i$ are shifted by small perturbative corrections). The collectivity results in an  energy shift that can be large even if the elementary interaction  $V$ is small.

I turn now to ultra-soft modes which exist in the region of very small momenta, i.e., when $p\ll g^2 T$. In this region, the term $k\cdot p\ll g^2T^2$ in the denominator of Eq.~(\ref{eq:sigma-mu}) becomes negligible compared to $\delta m^2\sim g^2T^2$, and the dispersion relation is given simply by $\Sigma(\omega,p)=0$.  Crucial to the existence of such a mode is therefore the fact that $\delta m^2$ is non vanishing, that is, the thermal asymptotic masses of the bosons and the fermions differ. This has been interpreted \cite{Lebedev:1989rz} as a spontaneous breaking of an approximate supersymmetry by finite temperature effects (those at the origin of the thermal masses). In analogy with what happens when continuous symmetry are spontaneously broken and Goldstone bosons appear,  one expects here a massless excitation to appear, with however fermionic quantum numbers: such excitation has been called ``goldstino''. Such modes have been studied in  simple supersymmetric models like the Wess-Zumino model\cite{Kratzert:2003cr}. There, because the supersymmetry algebra involves the energy-momentum tensor, it turns out that the goldstino becomes the analog of hydrodynamic sound modes, whose group velocity $v=1/3$ is entirely determined by ideal  thermodynamics. This group velocity is identical to that of the plasmino at small $p$;  it is also the same as that of  the goldstino of QED or QCD, or even that found in the simple scalar Yukawa model\cite{Hidaka:2011rz},  where the hydrodynamical analogy is less obvious. In order to understand better the nature of this goldstino, we have recently studied how it is affected by a chemical potential. A chemical potential breaks charge conjugation symmetry, as well as supersymmetry, in an explicit fashion. But if the breaking is not too strong, one finds that the goldstino still exists, and acquires a mass directly proportional to the explicit symmetry breaking term, again in complete analogy with what happens to usual Goldstone bosons \cite{Blaizot:2014hka}.

   \section{Cold atom systems}
   
    It turns out that one may soon have the possibility to  study some of these issues with cold atom systems (see, e.g., \cite{Yu:2007xb}, and references therein. Another related analogy between cold atoms and QCD matter can be found in \cite{Maeda:2009ev}). A possible experimental setup to do that was proposed in Ref.~\cite{Shi:2009ak}. It consists of two kinds of fermions (denoted $f$ and $F$) and their bound state, a boson $b$,  living on an optical lattice. The interactions are such that $f$ and $F$ form a bound state, but only $f$ interacts with $b$, while the interaction between $F$ and $b$ is negligible. The hamiltonian of the system is  written as $H=H_0+H_1$ with 
\beq
H_0=-t\sum_{<ij>}\left( f_i^\dagger f_j+b_i^\dagger b_j \right)-\mu_f \sum_i      f_i^\dagger f_i-\mu_b\sum_i b_i^\dagger b_i 
\eeq
and 
\beq
H_1=\frac{U_{\rm bb}}{2}\sum_i  b_i^\dagger b_i^\dagger b_i b_i +U_{\rm bf} \sum_i b_i^\dagger b_i f_i^\dagger f_i .
\eeq
Here     $\sum_{<ij>}$ denotes a sum over nearest neighbors, $\mu_{\rm f}$ and $\mu_{\rm b}$ are the chemical potentials of fermions and bosons, respectively.     

  Since we are interested in long wavelength phenomena, with typical momenta much smaller than the inverse lattice spacing, we shall move to continuum notation (in 2$d$ space). We introduce field operators for bosons,  $[b(\x),b^\dagger(\y) ]=\delta^{(2)}(\x-\y)$, and for fermions, $\{f(\x),f^\dagger(\y) \}=\delta^{(2)}(\x-\y)$, as well as number operators
   \beq
 N_{\rm b}=\int {\rm d}^2\x \,b^\dagger (\x)b(\x),\qquad    N_{\rm f}=\int {\rm d}^2\x \,f^\dagger (\x)f(\x). 
 \eeq
 Important in the present discussion is the so-called  ``supercharge''
 \beq
 Q\equiv \int {\rm d}^2\x \, q(\x),\qquad  q(\x)\equiv b(\x)f^\dagger(\x).
 \eeq
 The operator $Q$ transform a boson into a fermion while the operator $Q^\dagger$ does the opposite.
 This supercharge, together with the number operators, form the supersymmetric algebra
 \beq
 \{ Q,Q^\dagger \}=N,\quad [N,Q]=[N,Q^\dagger]=0,\quad [\Delta N,Q]=Q,
 \eeq
 where $N\equiv N_{\rm b}+N_{\rm f}$, and $\Delta N\equiv (N_{\rm f}-N_{\rm b})/2$.
 
 When the chemical potentials are equal, $\mu_{\rm f}=\mu_{\rm b}$,  and when the strength of the interactions are also equal, $U_{\rm bb}=U_{\rm bf}$, the hamiltonian $H$ is supersymmetric, that is,  $[H,Q]=0$, as easily verified by a direct calculation.
     
  Let me emphasize here  that the present supersymmetry should be viewed as a dynamical symmetry, analogous to that introduced in nuclear physics to describe complex nuclear spectra\cite{Iachello:1980av}.  It is not associated with space-time symmetries as it is in high energy physics.
     
     Supersymmetry can be broken. This can be done in two ways. An explicit breaking occurs when the chemical potential of the fermions differs from that of the bosons. Consider indeed the grand canonical hamiltonian
    \beq
    \hat H=H-\mu_{\rm f}N_{\rm f}-\mu_{\rm b}N_{\rm b}=H=\mu N-\Delta\mu\,\Delta N,
    \eeq
    where we have set $\mu\equiv (\mu_{\rm f}+\mu_{\rm b})/2$ and $\Delta\mu\equiv \mu_{\rm f}-\mu_{\rm b}$. Clearly, $[Q,\hat H]=\Delta \mu Q$, and the symmetry is explicitly broken when $\Delta\mu\ne 0$. 
     Supersymmetry can also be spontaneously broken by the presence of matter. The spontaneous breakdown of symmetry is accompanied by a corresponding Goldstone excitation,  the goldstino.
     All this is quite analogous to what we have just discussed in the context of the quark-gluon plasma. 
     
     \begin{figure}
\begin{center}
\includegraphics[scale=0.5]{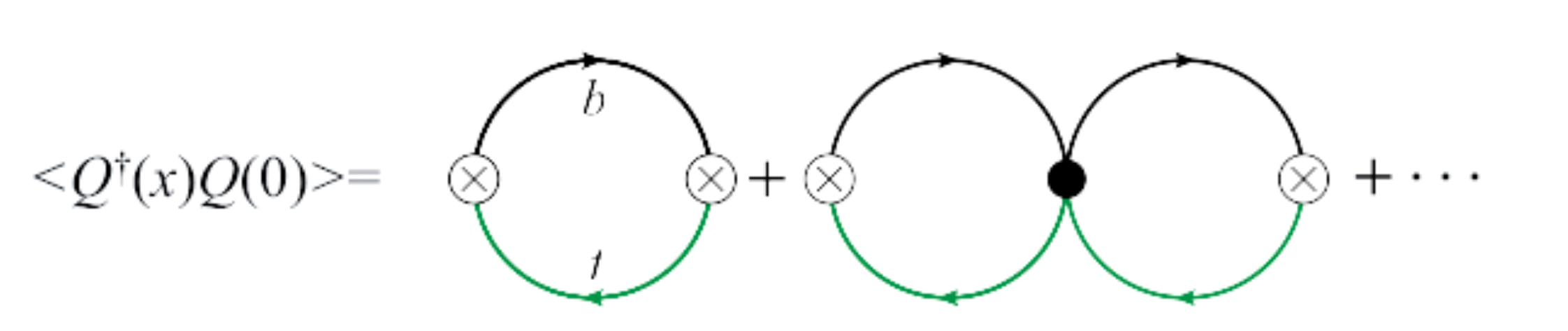}
\caption{ The sum of bubbles for the supercharge correlator  in the Random Phase Approximation.\label{RPA}}
\end{center}
\end{figure}

     In the present context the spectrum can also be obtained from simple approximations. Here the relevant approximation is the random phase approximation, which amounts to resum the chain of  boson-fermion bubble graphs  for the correlator of two supercharges. These graphs are displayed in Fig.~\ref{RPA}. One then gets the retarded response
     \beq
     G^R(p)\simeq -\frac{Z}{\omega-\Delta\mu +\alpha p^2},
     \eeq
     from which the spectrum of excitations is easily deduced. A massless mode appears when $\Delta\mu=0$: this is the goldstino, whose dispersion relation is here quadratic in momentum. A finite $\Delta\mu$ breaks the supersymmetry explicitly and  makes the goldstino massive. There is an analogy between the goldstino and the magnons in a ferromagnet. In both cases, we have conserved  charges. For the goldstino, the conserved charges are the super charges. In a ferromagnet, the conserved charges are the components of the local magnetization. These obey commutation relations. Broken symmetries emerge when the commutator of these conserved charges does not vanish. The dispersion relation is, in both cases, quadratic in the momentum. These issues, as well as their relations to the corresponding ones in the quark-gluon plasma, are being investigated and will be reported on in a forthcoming publicaion \cite{superBHD}.

c
My renewed interest in these issues owes much to my recent collaboration with D. Satow.
My research is supported by  the European Research Council under the
Advanced Investigator Grant ERC-AD-267258

\end{document}